\documentclass[letterpaper,12pt, onecolumn, nodraft]{IEEEtran}
\usepackage{graphicx,latexsym,epsfig,amssymb,amsmath,subfigure}
\usepackage{setspace}
\doublespace
\begin{document}

\title{A Novel Cyber-Insurance Model for Internet Security}


\author{Ranjan Pal, Leana Golubchik, and Konstantinos Psounis\\University of Southern California\\Email: \{rpal, leana, kpsounis\}@usc.edu}
\maketitle
\begin{abstract}
Internet users such as individuals and organizations are subject to different types of epidemic risks such as worms, viruses, and botnets. To reduce the probability of risk, an Internet user generally invests in self-defense mechanisms like antivirus and antispam software. However, such software does not completely eliminate risk. Recent works have considered the problem of residual risk elimination by proposing the idea of cyber-insurance. In reality, an Internet user faces risks due to security attacks as well as risks due to non-security related failures (e.g., reliability faults in the form of hardware crash, buffer overflow, etc.) . These risk types are often indistinguishable by a naive user. However, a cyber-insurance agency would most likely insure risks only due to security attacks. In this case, it becomes a challenge for an Internet user to choose the right type of cyber-insurance contract as standard optimal contracts, i.e., contracts under security attacks only, might prove to be sub-optimal for himself.

In this paper, we address the problem of analyzing cyber-insurance solutions when a user faces risks due to both, security as well as non-security related failures. We propose \emph{Aegis}, a novel cyber-insurance model in which the user accepts a fraction \emph{(strictly positive)} of loss recovery on himself and transfers rest of the loss recovery on the cyber-insurance agency. We mathematically show that given an option, Internet users would prefer Aegis contracts to traditional cyber-insurance contracts, under all premium types. This result firmly establishes the non-existence of traditional\footnote{Traditional cyber-insurance contracts are those which do not operate on non-security related losses in addition to security related losses, and do not give a user that option of being liable for a fraction of insurer advertised loss coverage. In Section \ref{sec-related} we cite recent important papers on traditional cyber-insurance and differentiate their work from our cyber-insurance model.} cyber-insurance markets when Aegis contracts are offered to users. We also derive an interesting counterintuitive result related to the Aegis framework: we show that an increase(decrease) in the premium of an Aegis contract \emph{may not} always lead to decrease(increase) in its user demand. In the process, we also state the conditions under which the latter trend and its converse emerge. Our work proposes a new model of cyber-insurance for Internet security that extends all previous related models by accounting for the extra dimension of non-insurable risks. Aegis also incentivizes Internet users to take up more personal responsibility for protecting their systems. 

\end{abstract}
%
\IEEEpeerreviewmaketitle
\section{Introduction} \label{sec-intro}
The Internet has become a fundamental and an integral part of our daily lives. Billions of people nowadays are using the Internet for various types of applications. However, all these applications are running on a network, that was built under assumptions, some of which are no longer valid for today's applications, e,g., that all users on the Internet can be trusted and that there are no malicious elements propagating in the Internet. On the contrary, the infrastructure, the users, and the services offered on the Internet today are all subject to a wide variety of risks. These risks include denial of service attacks, intrusions of various kinds, hacking, phishing, worms, viruses, spams, etc. In order to counter the threats posed by the risks, Internet users\footnote{The term `users' may refer to both, individuals and organizations.} have traditionally resorted to antivirus and anti-spam softwares, firewalls, and other add-ons to reduce the likelihood of being affected by threats. In practice, a large industry (companies like \emph{Symantec, McAfee,} etc.) as well as considerable research efforts are centered around developing and deploying tools and techniques to detect threats and anomalies in order to protect the Internet infrastructure and its users from the resulting negative impact.

In the past one and half decade, protection techniques from a variety of computer science fields such as cryptography, hardware engineering, and software engineering have continually made improvements. Inspite of such improvements, recent articles by Schneier \cite{sch} and Anderson \cite{ranr}\cite{amr} have stated that it is impossible to achieve a 100\% Internet security protection. The authors attribute this impossibility primarily to four reasons: 
\begin{itemize}
\item New viruses, worms, spams, and botnets evolve periodically at a rapid pace and as a result it is extremely difficult and expensive to design a security solution that is a panacea for all risks.
\item The Internet is a distributed system, where the system users have divergent security interests and incentives, leading to the problem of `misaligned incentives' amongst users. For example, a rational Internet user might well spend \$20 to stop a virus trashing its hard disk, but would hardly have any incentive to invest sufficient amounts in security solutions to prevent a service-denial attack on a wealthy corporation like an Amazon or a Microsoft \cite{varian}. Thus, the problem of misaligned incentives can be resolved only if liabilities are assigned to parties (users) that can best manage risk. 
\item The risks faced by Internet users are often correlated and interdependent. A user taking protective action in an Internet like distributed system creates positive externalities \cite{hh} for other networked users that in turn may discourage them from making appropriate security investments, leading to the `free-riding' problem \cite{gccr}\cite{jaw}\cite{mybm}\cite{oom}. 
\item Network externalities affect the adoption of technology. Katz and Shapiro \cite{kschr} have determined that externalities lead to the classic S-shaped adoption curve, according to which slow early adoption gives way to rapid deployment once the number of users reaches a critical mass. The initial deployment is subject to user benefits exceeding adoption costs, which occurs only if a minimum number of users adopt a technology; so everyone might wait for others to go first, and the technology never gets deployed. For example DNSSEC, and S-BGP are secure protocols that have been developed to better DNS and BGP in terms of security performance. However, the challenge is getting them deployed by providing sufficient internal benefits to adopting entities.
\end{itemize}
In view of the above mentioned inevitable barriers to 100\% risk mitigation, the need arises for alternative methods of risk management in the Internet. Anderson and Moore \cite{amr} state that microeconomics, game theory, and psychology will play as vital a role in effective risk management in the modern and future Internet, as did the mathematics of cryptography a quarter century ago. In this regard, \emph{cyber-insurance} is a psycho-economic-driven risk-management technique, where risks are transferred to a third party, i.e., an insurance company, in return for a fee, i.e., the \emph{insurance premium}. The concept of cyber-insurance is growing in importance amongst security engineers. The reason for this is three fold: (i) ideally, cyber-insurance increases Internet safety because the insured increases self-defense as a rational response to the reduction in insurance premium \cite{kmy1}\cite{kmy2}\cite{bs}\cite{yd}, a fact that has also been mathematically proven by the authors in \cite{leb3}\cite{leb}, (ii) in the IT industry, the mindset of `absolute protection' is slowly changing with the realization that absolute security is impossible and too expensive to even approach while adequate security is good enough to enable normal functions - the rest of the risk that cannot be mitigated can be transferred to a third party \cite{kmy3}, and (iii) cyber-insurance will lead to a market solution that will be aligned with economic incentives of cyber-insurers and users (individuals/organizations) - the cyber-insurers will earn profit from appropriately pricing premiums, whereas users will seek to hedge potential losses. In practice, users generally employ a simultaneous combination of retaining, mitigating, and insuring risks \cite{bs2}.

\emph{Research Motivation:} The concept of cyber-insurance as proposed in the security literature covers losses only due to security attacks. However, in reality, security losses are not the only form of losses. Non-security losses (e.g., reliability losses) form a major loss type, where a user suffers, either because of hardware malfunction due to a manufacturing defect or a software failure (e.g., buffer overflow caused by non-malicious programming or operational errors\footnote{A buffer overflow can also be caused by a malicious attack by hackers. Example of such attacks include the Morris worm, Slapper worm, and Blaster worm attacks on Windows PCs.})\cite{hs-security}. A naive Internet user would not be able to distinguish between a security or a non-security failure and might be at a disadvantage w.r.t. buying traditional cyber-insurance contracts. That is, on facing a risk, the user would not know whether the cause of the risk is a security attack or a non-security related failure\footnote{Irrespective of whether a loss due to a risk is because of a security attack or a non-security failure, the effects felt by a user are the same in both cases.}. The disadvantage is due to the fact that traditional cyber-insurance would only cover those losses due to security attacks, whereas an Internet user may incur a loss that occurs due to a non-security problem and not get covered for it\footnote{We assume here that the loss covering agency can distinguish between both types of losses and it does not find it suitable to cover losses due to hardware or software malfunctions, as it feels that they should be the responsibility of the hardware and software vendors (e.g., some computer service agencies in India employ experts who could distinguish between the two loss types, and these experts may be hired by the loss recovery agency also.).}. In such cases, it is an interesting problem to investigate the demand for traditional cyber-insurance as it seems logical to believe that an Internet user might not be in favor of transferring complete loss recovery liability to a cyber-insurer as the former would would have to pay the premium and at the same time bear the valuation of the loss on being affected by non-security related losses.  In this paper, we analyze the situation of Internet users buying cyber-insurance when they face risks that may arise due to non-security failures or security attacks. We propose an alternative model of cyber-insurance, i.e., Aegis, in this regard and show that given an option between cyber-insurance and Aegis contracts, an Internet user would \emph{always} prefer the latter. We make the following contributions in the paper. 
\begin{enumerate}
\item We propose a novel\footnote{Our cyber-insurance model is novel because we model partial insurance, whereas existing works related to traditional cyber-insurance model full and partial insurance coverage but not partial insurance. The notion of partial insurance can be explained as follows: in traditional cyber-insurance models, only the cyber-insurer has the say on the amount of coverage it would provide to its clients and in turn the premiums it would charge, whereas in the Aegis model, the clients get to decide on the fraction of the total amount of advertised insurance coverage it wants and in turn the proportional premiums it would pay, given an advertised contract. Thus, in traditional cyber-insurance, it is mandatory for users to accept the insurance policy in full, whereas in the Aegis model users have the option of accepting the insurance policy in partial.} model of cyber-insurance, Aegis, in which Internet users need not transfer the total loss recovery liability to a cyber-insurer, and may keep some liability to themselves, i.e., an Internet user may not transfer the entire risk to an insurance company. Thus, as an example, an Internet user may rest 80\% of his loss recovery liability to a cyber-insurer and may want to bear the remaining 20\% on his own. Our model captures the realistic scenario that Internet users could face risks from security attacks as well as from non-security related failures. It is based on the concept of co-insurance in the traditional insurance domain. (See Section \ref{sec-model}.)\\
\item We mathematically show that when Internet users are risk-averse, Aegis contracts are \emph{always} the user preferred policies when compared to traditional cyber-insurance contracts. In this regard, the latter result de-establishes a market for traditional cyber-insurance.
The availability of Aegis contracts also \emph{incentivizes} risk-averse Internet users to rest some loss coverage liability upon themselves rather than shifting it all to a cyber-insurer. (See Section \ref{sec-efficacy}.) \\
\item We mathematically show that a risk-averse Internet user would prefer cyber-insurance of some type (Aegis or traditional) \emph{only} if it is mandatory for him to buy some kind of insurance, given that he faces risks due to both, security as well as non-security failures.  (See Section \ref{sec-efficacy}.) \\ 
\item We mathematically show the following counterintuitive results: (i) an increase in the premium of an Aegis contract \emph{may not} always lead to a decrease in its user demand and (ii) a decrease in the premium of an Aegis contract may not always lead to an increase in its user demand. In the process, we also state the conditions under which these trends emerge. The conditions give a guideline to cyber-insurers on how to increase or decrease their premiums in order to increase user demands for cyber-insurance.  (See Section \ref{sec-analysis}.)

\end{enumerate}

 \section{The Aegis Cyber-Insurance Model} \label{sec-model}
We consider the scenario where an Internet user faces risks\footnote{A risk is defined as the chance that a user faces a certain amount of loss.}  that arise due to security attacks from worms, viruses, etc., as well as due to non-security related failures. One example of non-security related problems arises due to reliability faults. In a seminal paper \cite{hs-security}, the authors identified operational and programming errors, manufacturing problems of software and hardware vendors, and buffer overflow as some examples of system reliability faults, which have effects on Internet users that are identical to the effects when they are affected by certain security threats (e.g., buffer overflow due to a malicious attack). On facing the negative effects, an Internet user in general cannot distinguish between the loss type. In this paper, we assume that a loss occurs either due to a security attack or a non-security related failure and not both, i.e., a unit of damage cannot occur simultaneously due to a security and a non-security failure. For example, a file or a part of it that has been damaged by a security attack cannot be damaged by a non-security fault at the same time.

We assume that cyber-insurers\footnote{A cyber-insurer could be an ISP, a third-party agency, or the government.} offer Aegis contracts to their clients, Aegis contracts unlike traditional insurance contracts allow the user to rest some fraction of loss recovery liability upon itself. For example, if the value\footnote{In this paper, like in all of existing cyber-insurance literature, we assume that loss and coverage have the same scalar unit. In reality, this may not be true. As an example, losing a valuable file may not be compensated by replacing the same file. In return, monetary compensation may result. Considering appropriate units of loss and coverage is an area of future work. } of a loss incurred by an Internet user equals $L$, and the insurance coverage advertised by an insurer equals $L - d$, where $d\ge0$, an Aegis contract would allow its client to rest a fraction, $1 - \theta$, of the coverage on itself and the remaining $\theta$ part on the cyber-insurer, whereas a traditional contract would fix the value of $\theta$ to 1. 
Our concept of Aegis contracts are based on the theory of \emph{co-insurance} in general insurance literature. It is logical to believe that a user will not prefer a $1 -\theta$ value that is large as it would mean that it wants to rest a substantial loss recovery liability on itself, thereby diminishing the importance of buying cyber-insurance. We assume that the value of $\theta$ is fixed between the user and the cyber-insurer prior to contract operation.

Most of our analysis in the paper will revolve around the final wealth of a risk-averse Internet user who may be subject to risks due to both security attacks and non-security related failures. We have the following equation regarding its final wealth according to the Aegis concept:
\begin{equation}
W = w_{0} + v - L_{S} - L_{NS} + \theta(I(L_{S}) - P),
\end{equation}
where $W$ is a random variable representing the final wealth of a user, $w_{0} + v$ is his constant initial wealth, with $v$\footnote{We divide the fixed initial wealth of a user into two parts for modeling simplicity.} being the constant total value of the object subject to loss as a result of a security attack or a non-security attack, $L_{S}$ is a random variable denoting loss due to security attacks, $L_{NS}$ is the random variable denoting loss due to non security related failures, and$I(L_{S})$ is the cyber-insurance function that decides the amount of coverage to be provided in the event of a security-related loss, where $0\le I(L_{S}) \le L_{S}$. We assume that both $L_{S}$ and $L_{NS}$ lie in the interval $[0,v]$. As mentioned previously, a given amount of loss can be caused either by a security attack or a by a non-security fault and not by both. In this sense the loss types are \emph{not independent} but are \emph{negatively correlated}. $P$ is the premium\footnote{$P$ is the premium corresponding to a $\theta$ value of 1, where $\theta$ is the level of cyber-insurance liability opted by a user.} charged to users in insurable losses and is defined as $P = (1 + \lambda)E(I(L_{S}))$. $\lambda $ is the loading factor and is zero for fair premiums and greater than zero for unfair premiums. We define $\theta\,\epsilon\,[0,1]$ as the \emph{level of cyber-insurance liability} opted for by a user. For example, a value of $\theta = 0.6$, implies that the user transfers 60\% of its insurance coverage liability to the cyber-insurer and keeps the rest 40\% of the coverage liability on himself, where the insurance coverage could be either full or partial. We observe from Equation (1) that depending on the liability level, a user pays proportional premiums to the cyber-insurer.

We define the expected utility of final wealth\footnote{In economic and risk analyses, dealing with the expected utility of final wealth is a standard approach and it arises from the von Neumann-Morgenstern model of expected utility \cite{nmr}.}  of an Internet user as
\begin{equation}
E(W) = A + B + C + D,
\end{equation}
where
\[A = \int\int_{0 < L_{S} \le v, L_{NS} = 0} u(w_{0} + v - L_{S} - L_{NS} + \theta(I(L_{S}) - P))\cdot g(L_{S}, L_{NS})dL_{1}\cdot d{L_{NS}},\]
\[B = \int\int_{0 < L_{NS} \le V, L_{S} = 0} u(w_{0} + v - L_{S} - L_{NS} + \theta(I(L_{S}) - P))\cdot g(L_{S}, L_{NS})dL_{S}\cdot d{L_{NS}},\]
\[C = \int\int_{0 < L_{S}, 0 < L_{NS}} u(w_{0} + v - L_{S} - L_{NS} + \theta(I(L_{S}) - P))\cdot g(L_{S}, L_{NS})dL_{S}\cdot d{L_{NS}},\]
and
\[D = \beta\cdot u(w_{0} + v - \theta\cdot P),\]
with $A$, $B$, $C$, and $D$ being the components of expected utility of final wealth when there is a loss due to a security attack only, a non-security related failure only, a security attack as well as a non-security related failure, and no failure of any kind, respectively. $u$ is a twice continuously differentiable risk-averse concave utility function of wealth of a user. 

We define the joint probability density function, $g()$, of $L_{S}$ and $L_{NS}$ as
\begin{equation}
g(L_{S},L_{NS}) = \left\{
\begin{array}{rl}
\alpha\cdot f_{S}(L_{S}) & 0 < L_{S}\le v, L_{NS} = 0\\
(1 - \alpha - \beta)\cdot f_{NS}(L_{NS}) & 0 < L_{NS}\le v, L_{S} = 0,\\
0 & 0 < L_{S}\le v, 0 < L_{NS}\le v
\end{array}\right.
\end{equation}
where $\alpha$ is the probability\footnote{We plan to estimate $\alpha$ using correlation models.}  of loss occurring due to a security attack, and $\beta$ is the probability of no attack due to either a security or a non security attack. $f_{S}(L_{S})$ and $f_{NS}(L_{NS})$ are the univariate density functions of losses due to a security attack and non security attack respectively. The joint probability density function has three components: 1) the case where there is a loss only due to a security attack, 2) the case when there is a loss only due to a non-security related failure, and 3) the case when a loss occurs due to both types of risks.

Based on $g()$, Equation (1) can be re-written as
\begin{equation}
E(W) = A1 + B1 + C1,
\end{equation}
where
\[A1 = \int_{0}^{v}u(w_{0} + v - L_{S} + \theta(I(L_{S}) - P))\alpha\cdot f_{S}(L_{S})dL_{S},\]
\[B1 = \int_{0}^{v}u(v_{0} + v - L_{NS} - \theta(P))(1 - \alpha - \beta)\cdot f_{NS}(L_{NS})dL_{NS},\]
and
\[C1 = \beta\cdot u(w_{0} + v - \theta\cdot P),\]
with $A1$, $B1$, and $C1$ being the components of expected utility of final wealth when there is a loss due to a security attack only, a non-security related failure only, and no failure of any kind, respectively. 

In the following sections, we adopt the Aegis model of cyber-insurance and derive results in the form of theorems and propositions.  
\section{ Efficacy of Aegis Contracts} \label{sec-efficacy}
In this section, we investigate whether Aegis contracts are preferred by Internet users over traditional cyber-insurance contracts, and if yes, then under what conditions. In this regard, we state the following theorems that establish results regarding the user demand for Aegis contracts when compared to traditional cyber-insurance contracts.\\
\textbf{Theorem 1.} \emph{Risk-averse Internet users always prefer Aegis contracts to traditional cyber-insurance contracts when non-insurable losses exist, irrespective of whether the cyber-insurance premium charged in an Aegis contract is fair $(\lambda = 0)$ or unfair $(\lambda > 0)$}\footnote{The comparison is based on \emph{equal} degrees of fairness or unfairness between an Aegis contract and a traditional cyber-insurance contract.}.\\

\emph{Proof:} Taking the first derivative of $E(W)$ w.r.t. $\theta$, and equating it to zero, we get the first order condition as
\begin{equation}
\frac{dE(W)}{d\theta} = A2 + B2 + C2 = 0,
\end{equation}
where
\[A2 = \int_{0}^{v}u'(w_{0} + v - L_{S} + \theta(I_{L_{S}} - P))(I(L_{S}) - P)\alpha\cdot f_{S}(L_{S})dL_{S},\]
\[B2 = \int_{0}^{v}u'(w_{0} + v - L_{NS} - \theta(P))(-P)(1 - \alpha - \beta)\cdot f_{NS}(L_{NS})dL_{NS},\]
and
\[C2 =  \beta\cdot u'(w_{0} + v - \theta\cdot P)(-P).\]
Now substituting $I(L_{S}) = L_{S}$ ( indicating full coverage) and $\theta = 1$ (indicating no co-insurance) into the first order condition, we get
\begin{equation}
\frac{dE(W)}{d\theta} = A3 + B3 + C3 = 0,
\end{equation}
where
\[A3 = \int_{0}^{v}u'(w_{0} + v - P)(L_{S} - P)\alpha\cdot f_{S}(L_{S})dL_{S},\]
\[B3 = \int_{0}^{v}u'(w_{0} + v - L_{S} - P)(-P)(1 - \alpha - \beta)\cdot f_{NS}(L_{NS})dL_{NS},\]
and
\[C3 =  \beta\cdot u'(w_{0} + v - P)(-P).\]
Re-arranging the integrals we get
\[A3 = u'(w_{0} + v - P)\cdot \alpha\int_{0}^{v}(L_{S} - P) f_{S}(L_{S})dL_{S},\]
and
\[B3 = (-P)(1 - \alpha - \beta)\int_{0}^{v}u'(w_{0} + v - L_{NS} - P) f_{NS}(L_{NS})dL_{NS},\]
Now using the fact that $E(I(L_{S})) = \alpha\cdot\int_{0}^{v}L_{S}\cdot f_{S}(L_{S})dL_{S} = P$ (fair premiums), we have the following equation
\begin{equation}
\frac{dE(W)}{d\theta} = A4 + B4,
\end{equation}
where
\[A4 = u'(w_{0} + v - P)(1 - \alpha - \beta)P\]
and
\[B4 = (-P)(1 - \alpha - \beta)\int_{0}^{v}u'(w_{0} + v - L_{NS} - P) f_{NS}(L_{NS})dL_{NS}.\]
Since a user has a risk-averse utility function, we have $u'(w_{0} + v - L_{NS} - P) > u'(w_{0} + v - P)\, \forall L_{NS} > 0$. Thus, $\frac{dE(W)}{d\theta} < 0$ at $\theta = 1$. This indicates that the optimal value of $\theta$ is less than 1 for fair insurance premiums. On the other hand, even if we consider unfair premiums with a load factor $\lambda > 0$, we get $\frac{dE(W)}{d\theta} < 0$. Therefore in this case also the optimal value of $\theta$ is less than 1.  $\blacksquare$\\ \\
\emph{Implication of Theorem 1.} The theorem implies that risk-averse users would always choose Aegis cyber-insurance contracts over traditional cyber-insurance contracts, when given an option. \\
\emph{Intuition Behind Theorem 1.} In situations where a risk-averse user cannot distinguish between losses due to a security attack or a non-security failure, he would be conservative in his investments in insurance (as he could pay premiums and still not get covered due to a non-insurable loss) and would prefer to invest more in self-effort for taking care of his own system so as to minimize the chances of a loss. \emph{Thus, in a sense the Aegis model incentivizes risk-averse Internet users to invest more in taking care of their own systems than simply rest the entire coverage liability upon a cyber-insurer.}   \\
\textbf{Theorem 2. }\emph{When risks due to non-insurable losses are increased in a first order stochastic dominant\footnote{Let $X$ and $Y$ be two random variables representing risks. Then $X$ is said to be smaller than $Y$ in first order stochastic dominance, denoted as $X \le_{ST} Y$ if the inequality $VaR[X;p] \le VaR[Y; p]$ is satisfied for all $p\,\epsilon\,[0,1]$, where $VaR[X;p]$ is the value at risk and is equal to $F_{X}^{-1}(p)$. First order stochastic dominance implies dominance of higher orders. We adopt the stochastic dominant approach to comparing risks because a simple comparison between various moments of two distributions may not be enough for a correct prediction about the dominance of one distribution over another.} sense, the demand for traditional cyber-insurance amongst all risk-averse Internet users decreases.} \\

\emph{Proof.} Again consider the first order condition
\begin{equation}
\frac{dE(W)}{d\theta} = A2 + B2 + C2 = 0,
\end{equation}
where
\[A2 = \int_{0}^{v}u'(w_{0} + v - L_{S} + \theta(I_{L_{S}} - P))(I(L_{S}) - P)\alpha\cdot f_{S}(L_{S})dL_{S},\]
\[B2 = \int_{0}^{v}u'(w_{0} + v - L_{NS} - \theta(P))(-P)(1 - \alpha - \beta)\cdot f_{NS}(L_{NS})dL_{NS},\]
and
\[C2 =  \beta\cdot u'(w_{0} + v - \theta\cdot P)(-P).\]
We observe that when $L_{NS}$ is increased in a first order stochastic dominant sense\footnote{Let $X$ and $Y$ be two random variables representing risks. Then $X$ is said to be smaller than $Y$ in first order stochastic dominance, denoted as $X \le_{ST} Y$ if the inequality $VaR[X;p] \le VaR[Y; p]$ is satisfied for all $p\,\epsilon\,[0,1]$, where $VaR[X;p]$ is the value at risk and is equal to $F_{X}^{-1}(p)$. First order stochastic dominance implies dominance of higher orders. We adopt the stochastic dominant approach to comparing risks because a simple comparison between various moments of two distributions may not be enough for a correct prediction about the dominance of one distribution over another.} and $f_{S}(L_{S})$ and $\beta$ remain unchanged, the premium for insurance does not change. An increase in $L_{NS}$ in the first order stochastic dominant sense increases the magnitude of $\int_{0}^{v}u'(w_{0} + v - L_{NS} - \theta(P))(-P)(1 - \alpha - \beta)\cdot f_{NS}(L_{NS})dL_{NS}$, whenever $u'(w_{0} + v - L_{NS} - \theta(P))$ is increasing in $L_{NS}$. This happens when $u(W)$ is concave, which is the exactly the case in our definition of $u$. Thus, an increase in $L_{NS}$ in a first order stochastic dominant sense leads to the first order expression, $\frac{dE(W)}{d\theta}$, to become increasingly negative and results in reductions in $\theta$, implying the lowering of demand for cyber-insurance.  $\blacksquare$\\ \\
\emph{Implication of Theorem 2.} The theorem simply implies the intuitive fact that an increase in the risk due to non-insurable losses leads to a decrease in the demand of traditional cyber-insurance contracts, irrespective of the degree of risk-averseness of a user.\\ 
\emph{Intuition Behind Theorem 2.} The implication of Theorem 2 holds as the user would think that there are greater chances of it being affected by a loss and not being covered at the same time. An increase in the risk due to non-insurable losses also decreases the demand for Aegis contracts. However, according to Theorem 1, for the same amount of risk, Aegis contracts are preferred to traditional cyber-insurance contracts. \\ 
\textbf{Theorem 3.} \emph{When the risk due to non-insurable losses increases in the first order stochastic dominant sense, the expected utility of final wealth for any cyber-insurance contract (Aegis and traditional) falls when compared to the alternative of no cyber-insurance, for risk averse Internet users}. \\

\emph{Proof.} The expected utility of any cyber-insurance contract is given by the following
\begin{equation}
E(W) = A1 + B1 + C1,
\end{equation}
where
\[A1 = \int_{0}^{v}u(w_{0} + v - L_{S} + \theta(I(L_{S}) - P))\alpha\cdot f_{S}(L_{S})dL_{S},\]
\[B1 = \int_{0}^{v}u(w_{0} + v - L_{NS} - \theta(P))(1 - \alpha - \beta)\cdot f_{NS}(L_{NS})dL_{NS},\]
and
\[C1 = \beta\cdot u(w_{0} + v - \theta\cdot P).\]
When $\theta = 0$ (the case for no cyber-insurance), $E(W)$ reduces to
\begin{equation}
E(W) = A1' + B1' + C1',
\end{equation}
where
\[A1' = \int_{0}^{v}u(w_{0} + v - L_{S})\alpha\cdot f_{S}(L_{S})dL_{S},\]
\[B1' = \int_{0}^{v}u(w_{0} + v - L_{NS})(1 - \alpha - \beta)\cdot f_{2}(L_{NS})dL_{NS},\]
and
\[C1' = \beta\cdot u(w_{0} + v ).\]
Increases in $L_{NS}$ affect only the second terms in each of these utility expressions. Thus, we need to consider the change in the second order terms in the two utility expressions to observe the impact of the increase in $L_{NS}$. The difference in the second order terms is given as
\[\int_{0}^{v}u(w_{0} + v - L_{NS} - \theta(P))(1 - \alpha - \beta)\cdot f_{NS}(L_{NS})dL_{NS} - \int_{0}^{v}u(w_{0} + v - L_{NS})(1 - \alpha - \beta)\cdot f_{NS}(L_{NS})dL_{NS},\]
which evaluates to
\[\int_{0}^{v}[u(w_{0} + v - L_{NS} - \theta(P)) - u(w_{0} + v - L_{NS})](1 - \alpha - \beta)\cdot f_{NS}(L_{NS})dL_{NS},\]
where $[u(w_{0} + v - L_{NS} - \theta(P)) - u(w_{0} + v - L_{NS})]$ is decreasing in $L_{NS}$ under risk aversion and concave under user prudence. Thus, increases in $L_{NS}$ in the first order stochastic dominant sense reduces the expected utility of cyber-insurance relative to no cyber-insurance.  $\blacksquare$\\ \\
\emph{Implication of Theorem 3. } Theorem 3 provides us with an explanation of why risk-averse Internet users would be reluctant to buy cyber-insurance of any kind given an option between choosing and not choosing insurance, when risks due to non-security related losses are present along with risks due to security attacks.\\

\emph{Intuition Behind Theorem 3.} Theorem 3 holds because the expected utility to a risk-averse Internet user opting for a zero level of cyber-insurance liability is greater than that obtained when he opts for a positive level of cyber-insurance liability. 

Combining the results in theorems 1, 2, and 3, we conclude the following: 
\begin{itemize}
\item In the presence of non-insurable losses, the market for traditional cyber-insurance may not exist. 
\item When risk-averse Internet users have an option between traditional cyber-insurance, Aegis contracts, and no cyber-insurance, they may prefer the last option. Thus, Aegis contracts might be preferred by Internet users over traditional cyber-insurance contracts \emph{only} if it is mandatory for them to buy some kind of insurance. In general, Internet service providers (ISPs) might force its clients to sign up for some positive amount of cyber-insurance to ensure a more secure and robust Internet.   
\end{itemize}

\section{Sensitivity Analysis of User Demands} \label{sec-analysis}
In this section we conduct a sensitivity analysis of user demands for Aegis contracts. We investigate whether an increase in the premium charged by a contract results in an increase/decrease in user demand for the contract. The user demand is reflected in the $\theta$ value, i.e., user demand indicates the fraction of loss coverage liability a risk-averse user is willing to rest on the cyber-insurance agency. In an Aegis contract, to avoid insurance costs not related to a security attack, a risk-averse user takes up a fraction of loss coverage liability on himself  as it does not know beforehand whether he is affected by a security or a non security threat. Thus, intuitively, a decrease in a contract premium may not always lead to a user increasing his demand and analogously an increase in the premium may not always lead to a decrease in the user demands. The exact nature of the relationship between the premiums and user demand in this case depends on the degree of risk averseness of a user. To make the latter statement clear, consider an Internet user who is very risk averse. It would not matter to that user if there is a slight decrease in the premium amount because he might still not  transfer additional loss coverage liability to the cyber-insurer, given that he is unsure about whether the risk he faces is due to a security attack or a non security related issue. On the other hand a not so risk averse user may not decrease the amount of loss coverage liability rested upon a cyber-insurer, even if there is a slight increase in the cyber-insurance premiums. In this section we study the conditions under which there is an increase/decrease of user demand for Aegis contracts with change in contract premiums. We first provide the basic setup for sensitivity analysis, which is then followed by the study of the analysis results. 
\subsection{Analysis Setup}
Let a user's realized final wealth be represented as
\begin{equation}
W = w - L + \theta(L - P).
\end{equation}
Substituting $P = \lambda'E(L)$, we get
\begin{equation}
W = w - L + \theta(L - \lambda' E(L)),
\end{equation}
where $\lambda'$ equals $(1 + \lambda)$, $w$ is equal to $w_{0} + v$, $\theta$ lies in the interval $[0,1]$, $\lambda \ge 1$ is the gross loading factor of insurance, $L = L_{S} + L_{NS}$, and $\lambda E(L) = \alpha \int_{0}^{v}L\cdot f_{S}(L)dL$ is the premium payment for full insurance\footnote{By the term `full insurance', we imply a user resting its complete loss liability on the cyber-insurer, i.e., $\theta = 1$. Full insurance here does not indicate full insurance coverage.}  with $E$ being the expectation operator. The user is interested in maximizing his expected utility of final wealth in the von Neumann-Morgenstern expected utility sense and chooses a corresponding $\theta$ to achieve the purpose. Thus, we have the following optimization problem.
\[argmax_{\theta}E(U(W)) = E[U(w - L + \theta(L - \lambda' E(L))],\]
where $0 \le \theta \le 1$.
The first order condition for an optimum $\theta$ is given by
\begin{equation}
E'_{\theta}(U(W)) = E[U'(W)(L - \lambda' E(L))] = 0,
\end{equation}
which occurs at an optimal $\theta = \theta^{*}$. Integrating by parts the LHS of the first order condition and equating it to zero, we get
\begin{equation}
U'(W(0))\int_{0}^{v}(L - \lambda' E(L))dF_{S}(L) + \int_{0}^{v}U''(W(L))W'(L)\left(\int_{L}^{v}(t - \lambda' E(L))dF_{S}(t)\right) dL = 0,
\end{equation}
where $W(x)$ is the value of $W$ at $L = x$ and $W'(L) = -(1 - \theta) \le 0$.
The second order condition is given by
\begin{equation}
E''_{\theta}(U(W)) = E[U''(W)(L - \lambda' E(L))^{2}]  < 0,
\end{equation}
which is always satisfied for $U'' < 0$.  We now consider the following condition $C$, which we assume to hold for the rest of the paper.\\
\textbf{Condition C} - \emph{The utility function $U$ for a user is twice continuously differentiable, thrice piecewise  continuously differentiable\footnote{We consider the thrice piecewise continuously differentiable property of $U$ so that $A'(W)$ becomes piecewise continuous and is thus defined for all $W$.} and exhibits $U' > 0$, $U'' < 0$ with the coefficient of risk aversion, $A$, being bounded from above.}

The condition states the nature of the user utility function $U$, which is in accordance with the standard user utility function used in the insurance literature, with the additional restriction of thrice piecewise continuous differentiability of $U$ to make the coefficient of risk aversion well-defined for all $W$. We adopt the standard \emph{Arrow-Pratt} risk aversion measure \cite{mwg}, according to which the coefficient of risk aversion is expressed as (i) $A$ = $A(W) = -\frac{U''(W)}{U'(W)}$ for an \emph{absolute} risk averse measure and (ii) $R$ = $R(W) = -\frac{WU''(W)}{U'(W)}$  for a \emph{relative} risk averse measure. 
\subsection{Sensitivity Analysis Study}
In this section we study the change in user demands for Aegis contracts with variations in cyber-insurance premiums, under two standard risk-averse measures: (1) the decreasing absolute risk averse measure and (2) the decreasing relative risk averse measure. The term `decreasing' in both the risk measures implies that the risk averse mentality of users decrease with the increase in their wealth, which is intuitive from a user perspective. We are interested in investigating the sign of the quantity, $\frac{d\theta^{*}}{d\lambda'}$. The nature of the sign drives the conditions for an Aegis contract to be either more or less preferred by Internet users when there is an increase in the premiums, i.e., if $\frac{d\theta^{*}}{d\lambda'} \le 0$, an increase in cyber-insurance premium implies decrease in user demand, and $\frac{d\theta^{*}}{d\lambda'}  \ge 0$ implies an increase in user demand with increase in premiums.   

We have the following theorem and its corresponding proposition related to the conditions under which Internet users increase or decrease their demands for Aegis contracts, when the users are risk averse in an \emph{absolute} sense. \\ 
\textbf{Theorem 4.} \emph{For any arbitrary} $w$, $\lambda'$, $F$, \emph{and any} $U$ \emph{satisfying condition} $C$, \emph{(i)} $\frac{d\theta^{*}}{d\lambda'} \ge 0$ \emph{if and only if there exists} $\rho\,\epsilon\,\mathbb{R}$ \emph{such that}
\begin{equation}
\int_{L}^{w}[A(W(x))\theta^{*}(x - \lambda' E(L)) - 1]dF(x) \ge \rho\int_{L}^{w}\theta^{*}(x - \lambda' E(L))dF(x),
\end{equation}
\emph{and} \emph{(ii)} $\frac{d\theta^{*}}{d\lambda'} < 0$ \emph{if and only if there exists} $\rho\,\epsilon\,\mathbb{R}$ \emph{such that}
\begin{equation}
\int_{L}^{w}[A(W(x))\theta^{*}(x - \lambda' E(L)) - 1]dF(x) < \rho\int_{L}^{w}\theta^{*}(x - \lambda' E(L))dF(x),
\end{equation}
\emph{where} $L\,\epsilon\,[0,w]$ \emph{and} $F(\cdot)$ \emph{is the distribution function of loss} $L$. 
\\

\emph{Proof.} We know that $\frac{d\theta^{*}}{d\lambda'} = -\frac{E'_{\theta\lambda'}}{E''_{\theta}}$. Now $\frac{d\theta^{*}}{d\lambda} \le 0$ if and only if the following relationship holds because $E''_{\theta} < 0$. 
\begin{equation}
E'_{\theta\lambda'}(U(W(L))) = E[-U''(W(L))\theta^{*}E(L)(L - \lambda'E(L)) - U'(W(L))E(L)] \le 0
\end{equation}
or 
\begin{equation}
E\left\{\left(A(W(L)) - \frac{1}{\theta(L - \lambda'E(L))}\right)U'(W(L))(L - \lambda'E(L)\right\} \le 0
\end{equation}
The LHS of Equation 19 can be expressed via integration by parts as 
\[\int_{0}^{w}[A(W(L))\theta^{*}(L - \lambda'E(L)) - 1]U'(W(L))dF(L)\]
which evaluates to 
\[X + Y,\]
where
\[X = U'(W(0))\int_{0}^{w}[A(W(L))\theta^{*}(L - \lambda'E(L)) - 1]dF(L)\]
and 
\[Y = \int_{0}^{w}U''(W(L))W'(L)\left\{\int_{L}^{w}[A(W(t))\theta^{*}(x - \lambda'E(L)) - 1]dF(x)\right\} dL.\]
Now $X + Y \ge M + N$, where
\[M = U'(W(0))\int_{0}^{w}\rho(L - \lambda'E(L))dF(L)\]
and
\[N = \int_{0}^{w}U''(W(L))W'(L)\cdot \left (\rho\int_{L}^{w}(x - \lambda'E(L))dF(x)\right) dL.\]
Thus, $\frac{d\theta^{*}}{d\lambda'} \ge 0$, and the sufficient condition is proved. The proof of the necessary condition follows from Proposition 1' in \cite{ah}. Reversing Equation 16 we get the necessary and sufficient conditions for $\frac{d\theta^{*}}{d\lambda'} \le 0$, which is condition (ii) in Theorem 4. $\blacksquare$

\textbf{Proposition 1.}  \emph{There exists a} $\rho\,\epsilon\,\mathbb{R} - \{0\}$ \emph{such that Theorem 4 holds if the following two conditions are satisfied.}
\begin{equation}
\frac{(1 - \theta^{*})A'}{A} \le \theta^{*}
\end{equation}
and 
\begin{equation}
\int_{0}^{w}A(W(L))\left\{L - \lambda' E(L) - \frac{1}{\theta^{*}A(W(L))}\right\}dF(L) > 0
\end{equation}
\emph{Note on Theorem 4 and Proposition 1.} Theorem 4 and Proposition 1 are related to each other in the sense that Theorem 4 provides the necessary and sufficient conditions under which Internet users increase/decrease demands of Aegis contracts. The intuition behind the result in Theorem 4 is based on expected utility comparisons. For an increase in the $\lambda$ value, the expected utilities of a user are compared with and without a corresponding increase in $\theta$ value. We say that user demands for Aegis contracts increase (decrease) if there is an increase (decrease) in expected utility with an increase in the $\theta$ value, and we find the conditions for such situations to arise. Proposition 1 states that Theorem 4 always holds provided certain conditions are met. 
 
We have the following theorem that states the conditions under which Internet users increase or decrease their demands for Aegis contracts, when the users are risk averse in a \emph{relative} sense. \\ 
\textbf{Theorem 5.} \emph{For any arbitrary} $w$, $\lambda'$, $F$, \emph{and any} $U$ \emph{satisfying condition} $C$, \emph{(i)} $\frac{d\theta^{*}}{d\lambda'} \ge 0$ \emph{only if} $R(W) > 1$ and  \emph{(ii)} $\frac{d\theta^{*}}{d\lambda'} < 0$ \emph{only if} $R(W) \le 1$, where $W\,\epsilon\,[W(w), W(0)]$. \\

\emph{Proof:} We can rewrite Equation 16 as follows
\begin{equation}
\int_{L}^{w}\{\theta^{*}[A(W(x)) - \rho](x - \lambda'E(L)) - 1\}dF(x) \ge 0,
\end{equation}
which can be further rewritten as 
\begin{equation}
\int_{L}^{w}\{(R(W(x)) - 1) - A(W(x))(w_{0} - x) - \rho(x - \lambda'E(L))\}dF(x) \ge 0.
\end{equation}
The integral in Equation 23 is non-negative for all $L\,\epsilon\,[0,w]$ only if $R(W) > 1$ for some $W$. To see this it suffices to realize that $-A(W(L))(w_{0} - L) < 0$ for all $L\,\epsilon\,[0,w)$ as $L \le w_{0}$ and there exists $L\,\epsilon\,[0,w]$ at which $-\int_{L}^{w}\rho(L - \lambda'E(L))dF(x) < 0$ as $\int_{L}^{w}(x - \lambda'E(L))dF(x))$ alternates in sign on $(0,w)$. Now suppose by contradiction that $R(W) \le 1$ for all $W$. Substituting this into Equation 23 violates the condition stated in Equation 22 for some $L\,\epsilon\,[0,w]$.  Again by Theorem 4, we have that there exists utility function $U$ satisfying condition $C$ such that  $\frac{d\theta^{*}}{d\lambda'} \ge 0$ - a contradiction. Since $F$ is arbitrary, the result (i) in the theorem follows. By reversing the sign of the condition on $R(W)$ the result (ii) in the theorem follows. $\blacksquare$. 

\emph{Implication of Theorem 5.} The theorem implies that above a certain level of the degree of relative risk averseness, a user prefers Aegis contracts even if there is an increase in contract premiums. \\
\emph{Intuition Behind Theorem 5.} The coefficient of relative risk aversion is measured relative to the wealth of a user and thus more his wealth, lesser would be his concerns about losing money due to paying more cyber-insurance premiums, and not getting coverage on being affected by a non-security failure. The intuition is similar for the case when below a certain threshold of relative risk averseness, users reduce their demand for Aegis contracts. 

\section{Related Work}	\label{sec-related}
The field of cyber-insurance in networked environments has been fueled
by recent results on the amount of individual user self-defense investments
in the presence of network externalities\footnote{
	An externality  is a positive (external benefits) or negative (external costs) impact on an user not directly involved in an economic transaction.}.
The authors in \cite{gccr}\cite{jaw}\cite{leb5}\cite{leb4}\cite{mybm}\cite{oom} mathematically show
that Internet users invest too little in self-defense mechanisms relative
to the socially efficient level, due to the presence of network externalities.
These works highlight the role of positive externalities in preventing
users from investing optimally in self-defense. Thus, one challenge to
improving overall network security lies in incentivizing end-users to
invest in a sufficient amount of self-defense in spite of the
positive externalities they experience from other users in the network.
In response to this challenge, the works in \cite{leb5}\cite{leb4} modeled
network externalities and showed that a tipping phenomenon is possible,
i.e., in a situation where the level of self-defense is low, if a certain
fraction of population decides to invest in self-defense mechanisms,
then a large cascade of adoption in security features could be triggered,
thereby strengthening the overall Internet security. However, these
works did not state how the tipping phenomenon could be realized in
practice. In a series of recent works \cite{leb3}\cite{leb}, Lelarge and Bolot
have stated that under conditions of
no \emph{information asymmetry} \cite{wik}\cite{hv} between the insurer and
the insured, cyber-insurance \emph{incentivizes} Internet user investments
in self-defense mechanisms, thereby paving the path to triggering a cascade
of adoption. They also showed that investments in both self-defense mechanisms
and insurance schemes are quite inter-related in maintaining a socially
efficient level of security on the Internet. In a follow up work on joint self-defense and cyber-insurance investments, the authors in \cite{pg} show that Internet users invest more efficiently in self-defense investments in a cooperative environment when compared to a non-cooperative one, in relation to achieving a socially efficient level of security on the Internet. 

In spite of Lelarge and Bolot highlighting the role of cyber-insurance for
networked environments in incentivizing increasing of user security investments,
it is common knowledge that the market for cyber-insurance has not yet
blossomed with respect to its promised potential. Most recent works \cite{rabohme} \cite{ssfw} have
attributed this to
(1) \emph{interdependent security} (i.e., the effects of security investments
of a user on the security of other network users connected to it),
(2) \emph{correlated risk} (i.e., the risk faced by a user due to risks
faced by other network users),
and (3) \emph{information asymmetries} (i.e., the asymmetry between the insurer
and the insured due to one having some specific information about its risks
that the other does not have). In a recent work \cite{plg}, the authors have designed mechanisms to overcome the market existence problem due to information asymmetry, and show that a market for cyber-insurance exists in a single cyber-insurer setting.

However, none of the above mentioned works related to cyber-insurance address the scenario where a user faces risks due to security attacks as well as due to non-security related failures. The works consider attacks that occur due to security lapses only. In reality, an Internet user faces both types of risks and cannot distinguish between the types that caused a loss. Under such scenarios, it is not obvious that users would want to rest the full loss recovery liability to a cyber-insurer. We address the case when an Internet user faces risks due to both security as well as non-security problems, and show that users always prefer to rest some liability upon themselves, thus de-establishing the market for traditional cyber-insurance.
However, the Aegis framework being a type of a cyber-insurance framework also faces problems identical to the traditional cyber-insurance framework, viz., that of interdependent security, correlated risk, and information asymmetry. 

\section{Conclusion} \label{sec-conslusion}
In this paper we proposed Aegis, a novel cyber-insurance model in which an Internet user accepts a fraction (strictly positive) of loss recovery on himself and transfers the rest of the loss recovery on the cyber-insurance agency. Our model is specifically suited to situations when a user  cannot distinguish between similar types of losses that arise due to either a security attack or a non-security related failure. We showed that given an option, Internet users would prefer Aegis contracts to traditional cyber-insurance contracts, under all premium types. The latter result firmly establishes the non-existence of traditional cyber-insurance markets when Aegis contracts are offered to users. Furthermore, the Aegis model incentivizes risk-averse Internet users to invest more in taking care of their own systems than simply rest the entire coverage liability upon a cyber-insurer. We also derived two interesting counterintuitive results related to the Aegis framework, i.e., we showed that an increase (decrease) in the premium of an Aegis contract \emph{may not} always lead to a decrease (increase) in its user demand. As part of future work, we plan to investigate adverse selection and information asymmetry issues in Aegis contracts. 

\newpage
\bibliography{alluvion}
\bibliographystyle{plain}


\end{document}